\documentclass[aps,prl,nolongbibliography,notitlepage,twocolumn,superscriptaddress]{revtex4-2}

\usepackage{booktabs}
\usepackage{lipsum}
\usepackage{graphicx}
\usepackage{amsmath,amssymb,amsthm}
\usepackage{braket}
\usepackage{mathtools}
\usepackage{hyperref}
\usepackage{color}
\usepackage{float}
\usepackage{hyperref}
\usepackage[normalem]{ulem}
\usepackage{colonequals}
\usepackage{siunitx}
\usepackage{enumitem}
\usepackage{bbm}
\usepackage[dvipsnames]{xcolor}
\usepackage{tabularx}
\usepackage{mathrsfs}

\newcommand{\caron}[1]{\text{$\check{#1}$}}

\newcommand{\mf}{\text{mf}}
\renewcommand{\sp}{\text{sp}}

\renewcommand{\v}[1]{\boldsymbol{#1}}
\DeclareMathOperator{\Tr}{Tr}
\DeclareMathOperator{\tr}{tr}

\newcommand{\T}{\mathcal{T}}

\newcommand{\U}{\mathrm{U}}
\renewcommand{\H}{\mathcal{H}}

\newcommand{\br}{{\v r}}

\newcommand{\bR}{{\v R}}
\newcommand{\bk}{{\v k}}

\newcommand{\bq}{{\v q}}

\newcommand{\bG}{{\v G}}

\newcommand{\cG}{{\mathcal G}}

\newcommand{\ov}{\overline}

\DeclarePairedDelimiter\abs{\lvert}{\rvert}%

\begin{document}

\title{Exotic Carriers from Concentrated Topology: Dirac Trions as the Origin of the Missing Spectral Weight in Twisted Bilayer Graphene}
\author{Patrick J. Ledwith}
\author{Ashvin Vishwanath}
\author{Eslam Khalaf}
\affiliation{Department of Physics, Harvard University, Cambridge, MA 02138, USA}

\date{\today}
\begin{abstract}

The nature of charge carriers in twisted bilayer graphene (TBG) near $\nu = -2$, where superconductivity emerges, remains mysterious. While various symmetry-broken ground states have been proposed, experimental evidence of significant entropy persisting to low temperatures\cite{IlaniEntropy,saitoIsospinPomeranchukEffect2021,zhangHeavyFermionsMass2025} suggests that the disordered thermal state is a more natural starting point for understanding the normal state physics. Our previous work proposed that this thermal state in TBG hosts nearly decoupled \emph{nonlocal} moments and an exotic Mott semimetal at charge neutrality. This evolves into a spectrally imbalanced Mott state at non-zero integer fillings. Notably, at $\nu = -2$, the quasiparticle residue vanishes at the top of the valence band—precisely where superconductivity develops. 
The vanishing quasiparticle residue naturally leads to the following question: Which excitation accounts for the missing spectral weight, and how are they related  to electrons?  In this work, we demonstrate that the missing spectral weight corresponds to a trion excitation, which we explicitly construct and characterize. Remarkably, despite being composed of heavy particles away from the Gamma point, this trion is unexpectedly light. At charge neutrality, the electron and trion hybridize with a momentum-dependent phase that winds around zero, forming a massless Dirac cone. At finite doping $\nu < 0$, the Dirac cone acquires a mass with the valence (conduction) band becoming trion (electron)-like near the $\Gamma$ point. Finally, we show that such trion excitations evolve into the quasiparticles of the  intervalley Kekulé spiral (IKS) state at finite doping, where they represent electrons dressed by the IKS order parameter. More broadly, our work highlights the unusual spectral properties and excitations that emerge when fermions interact within topological bands with concentrated quantum geometry and topology. 
\end{abstract}
\maketitle

\emph{\bf Introduction}--- Twisted bilayer graphene (TBG) has been established as an emblematic platform for strongly interacting electrons in moiré systems, displaying a remarkably rich array of interacting phases including correlated insulators \cite{PabloMott,Dean-Young,efetov,EfetovScreening,PabloNematic,LiVafekscreening}, unconventional superconductivity \cite{PabloSC,Dean-Young,efetov,EfetovScreening,YoungScreening,NadjPergeSC,LiVafekscreening}, and quantum anomalous Hall states \cite{CorrelatedChernYoung, CorrelatedChernYazdani, CorrelatedChernNadjPerge, CorrelatedChernEvaAndrei, Pierce2021,VafekCorrelatedChern,Xie2021fractional,YoungQAH, DavidGGQAH}. Among these, the superconductor realized upon hole doping the correlated insulator at $\nu = -2$, and the associated normal state from which it emerges, remain one of the most intensely studied. The normal state exhibits strange metal transport \cite{PabloLinearT, YoungLinearT,EfetovSM} and signs of pseudogap physics \cite{jiangChargeOrderBroken2019,ohEvidenceUnconventionalSuperconductivity2021,choiInteractiondrivenBandFlattening2021}, while the superconductor exhibits evidence of nodal quasiparticles \cite{ohEvidenceUnconventionalSuperconductivity2021,banerjeeSuperfluidStiffnessTwisted2024} and a $T_c$ limited by phase fluctuations \cite{banerjeeSuperfluidStiffnessTwisted2024}. A key step in understanding these remarkable phenomena is the identification of the doped charge carriers and their interactions.  

At zero temperature, there is by now good evidence that the correlated states at non-zero integer fillings are generalized flavor ferromagnets at low strain \cite{BultinckHidden, Repellin19,TBGIVGroundState, hofmannFermionicMonteCarlo2022, parkerStrainInducedQuantumPhase2021, KwanKekule, Nuckolls_2023} and nematic semimetals \cite{huderElectronicSpectrumTwisted2018,MespleHeterostrain,bi_designing_2019,DaiEffectHeterostrain,WangUnusualMagnetotransport,STMNadjPerge,bocarsly2024imagingcoulombinteractionsmigrating,liuNematicTopologicalSemimetal2021,parkerStrainInducedQuantumPhase2021,soejimaEfficientSimulationMoire2020} or incommensurate spirals at intermediate strain \cite{KwanKekule, parkerStrainInducedQuantumPhase2021,Nuckolls_2023,WangIKSDMRG} relevant to most experiments. The cheapest charge excitations on doping such states away from neutrality are light single-particle excitations that can be captured using self-consistent Hartree-Fock \cite{BultinckHidden, TBGIVGroundState, KwanKekule, kangCascadesLightHeavy2021}. However, entropy measurements indicate the ordering temperature of such states is very small, of the order of a few Kelvin \cite{IlaniEntropy, YoungEntropy, zhangHeavyFermionsMass2025}. These findings suggest that the thermally disordered state is the natural starting point for understanding the physics near the correlated insulators, particularly at finite doping.

Recently, the authors have studied a model for TBG flat bands that helped shed light on the nature of the thermal state in an analytically controlled manner \cite{ledwith2024nonlocal}. Working in the projected Hilbert space of the flat bands, the authors employed a decomposition of the bands into a Chern basis of $\pm 1$ Chern bands \cite{BultinckHidden, TBGIVGroundState,ledwith2021strong}, which is a convenient basis to study interactions when the single particle dispersion is sufficiently small (close to the magic angle). A key observation of Ref.~\cite{ledwith2024nonlocal} is that the wavefunctions for a Chern band with concentrated charge density in real space also exhibit a concentrated Berry curvature with width denoted by a parameter $s \ll 1$. Such wavefunctions admit a universal form that depends on the small parameter $s$. For TBG, these correspond to flat bands where charge density is concentrated at the AA sites and Berry curvature is concentrated at the $\Gamma$ point. These wavefunctions were shown to closely match those of the Bistritzer-Macdonald (BM) model \cite{BMpaper} for a specific choice of $s \approx 0.25$ \cite{ledwith2024nonlocal}. This description of the flat bands was employed as the basis of an analytical treatment in which the nontrivial topology of the bands is only effective in a small $\sim s^2$ phase space. This enabled the calculation of several quantities of interest, including the single-particle Green's function and the associated spectral function in the thermally disordered state $T \gg Us^2$ at all integer fillings, where $U$ is the Hubbard interaction scale.

The spectral functions in the thermal state revealed an exotic and unusual behavior. At charge neutrality, the thermally disordered Mott state is a semimetal in which the electron spectral function describes a \emph{single} Dirac cone per spin per valley \cite{ledwith2024nonlocal}. This is inconsistent with any non-interacting or mean-field semimetal where symmetry and topology imply the existence of \emph{two} Dirac cones per spin per valley, suggesting that the thermally disordered state has some missing spectral weight corresponding to a second Dirac cone. The missing weight is even more manifest at non-zero integer fillings. Here, the Mott bands, now separated by a spectral gap, have a $\bk$-dependent quasiparticle residue that vanishes smoothly as $\bk$ approaches $\Gamma$ in the lower (upper) band for $\nu < 0$ ($\nu > 0$). This missing spectral weight suggests that the $\Gamma$ point band edge corresponds to an excitation orthogonal to any electron. Such excitations only hybridizes with the electron away from $\Gamma$ leading to a gradual increase in the spectral weight. Recent experiments using the quantum twisting microscope found indications of such missing weight near $\Gamma$ \cite{shahalnewexpt}. Interestingly, the missing spectral weight coincides with the doping regime where superconductivity and strange metal behavior are observed. This raises a key question: What is the precise nature of the quasiparticle excitation responsible for this missing spectral weight?

In this work, we explicitly construct this excitation as a two-electron one-hole ``trion" bound state. We find that its binding is kinetic in origin; the reduction in energy comes from a Dirac hybridization with mobile electrons. For the same reason, the trion is light --- it forms a massless cone with the electron at neutrality, and has a smaller mass than the electron away from neutrality. This is especially remarkable since the constituent electrons and holes of the trion come from the heavy, AA-localized, part of the band. Furthermore, since the trion excitation precisely describes the $\nu=-2$ valence band edge, it is the excitation that corresponds to the doped charge of the superconductor and exotic normal state. In fact, recent work \cite{zhao2025ancilla} showed that the aforementioned Mott states and certain exotic responses to perturbations can  be described within the ``ancilla" parton theory \cite{zhangPseudogapMetalFermi2020,zhouVariationalWavefunctionMott2024} and it was also suggested that the missing excitation could have a three-body nature. \\

\begin{figure*}
    \centering    
    \includegraphics{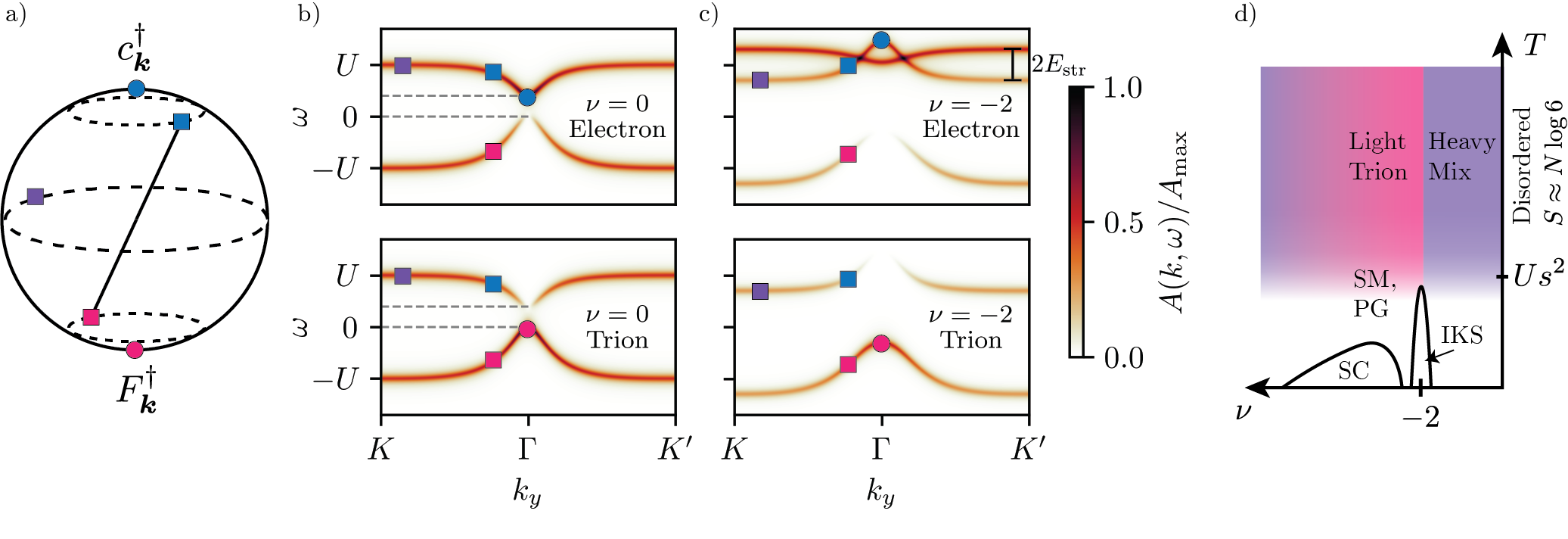}
    \caption{(a) Schematic illustration of the Bloch sphere describing a spinor with electron (north pole, blue circle) and trion (south pole, red cirle) states with different selected points indicated by the red, blue, purple squares. (b) Electron and trion spectral functions in the thermally disorder Mott semimetal state, illustrating the Dirac nature of the dispersion. A particle-hole breaking dispersion has been added, which acts as a mass term in the electron-trion space. (c) Electron and trion spectral function at $\nu = -2$ for $U s^2 \ll T \ll U$ including single-particle dispersion and strain ($T \ll E_{\rm str}$, $E_{\rm str} = 0.3 U$, $E_{\rm BM} = 0.2 U$). The top of the valence band where charge carriers are doped at $\nu = -2 - \delta$ has missing spectral weight in the electron Green's function and enhanced spectral weight in the trion Green's function. At the band edge, the excitations are purely trionic. (d) Schematic illustration of the TBG phase diagram close to $\nu = -2$ illustrating the nature of the charge carriers in the normal state above the superconducting dome. The carriers are light trions with small effective mass. On the side closer to charge neutrality, a heavy quasiparticle that is a mixture of electron and trion is the dominant charge carrier}.
    \label{fig:blochsphere_trions}
\end{figure*}

\emph{\bf Summary of results}--- 
The emergence of trion excitations can be motivated by Hubbard's original solution of his model's atomic limit \cite{hubbard1963electron}. From the commutator $[H_U, c_{\bR \alpha}] = -U F_{\bR \alpha}$, where $H_U = \sum_{\bR} U \delta n_\bR \delta n_\bR$, we see that the four-fermion Hubbard interaction tunnels a hole into the three-particle operator $F_{\bR \alpha} = \{c_{\bR \alpha}, \delta n_\bR \}$. Here $\delta n_\bR$ is the occupation relative to charge neutrality and $\alpha$ is a flavor index \footnote{We have expressed $H_U$ and $F$ differently than Hubbard so that particle-hole symmetry and generalization to $>2$ flavors is more manifest. We also use a Hubbard $U$ definition that is half the usual one to avoid dividing by two in every expression that it appears in within this manuscript.}.
Furthermore, when creating or annihilating a single particle excitation relative to the $U\gg T>0$ thermal state at charge neutrality, one can show that $[H_U, F_\bR]$ acts equivalently to $-Uc_\bR$. In fact, within this subspace of single particle states, $H_U$ acts as $-U \sum_\bR (c^\dag_{\bR \alpha} F_{\bR \alpha} + c_{\bR \alpha} F^\dag_{\bR \alpha}) $ and the fermions $c_{\bR},F_\bR$ act canonical due to $\langle \{c^\dag_\bR, F_\bR \}\rangle = 0$. The hole-like excitations of the Hubbard model are then electron-trion hybrids, $\frac{1}{\sqrt{2}}(c_\bR -  F_\bR)$, while the electron-like excitations are $\frac{1}{\sqrt{2}}(c^\dag_\bR +  F^\dag_\bR)$. 
With this context, it appears plausible that a Mott phase in a topological band could exhibit electron-trion hybridization that winds with $\bk$ as in a Dirac cone. In this paper, we show that this occurs in the Chern bands of TBG.

The first step is to define the trion operator in the topological setting. Our choice is motivated by the following considerations. Away from $\Gamma$, it should not be distinguished by any quantum number relative to the electron to allow mixing between the two. It should also reduce to the Hubbard trion since effects of topology are concentrated near $\Gamma$. At $\Gamma$ it should be protected from hybridizing with the electron, e.g. through a distinct $C_3$ eigenvalue. Together with the Hubbard context, these considerations suggest the following ansatz for the trion 
 \begin{equation}
  F_{\bk \alpha} =  \frac{1}{\sqrt{N}} \sum_\bR e^{i \bk \cdot \bR} \{c_{\bR \alpha}, \delta n_\bR \} - [A_\bk c_\bk]_\alpha,
  \label{eq:trioncreationtwo}
  \end{equation}
where $c_{\bR \gamma}  = \frac{1}{\sqrt{N}}\sum_\bk \chi_{\bk\gamma} c_{\bk \gamma} e^{-i \bk \cdot \bR}$, $\delta n_\bR = c^\dag_\bR c_\bR -4$, $N$ is the number of moiré lattice sites, $\gamma = \pm$ labels the $C=\pm$ Chern sector, and $[A_\bk]_{\alpha \beta}$ is a matrix in the band indices $\alpha,\beta$ chosen such that $\langle \{F_\bk, c^\dag_\bk \} \rangle = 0$, its explicit expression will be given later (for a strain-free symmetric charge neutrality state, $A = 0$). In a topological band, it is impossible to choose a gauge such that $c_{\bk,\gamma}$ is both smooth and periodic. The function $\chi_{\bk \gamma}$ is chosen such that the combination $\chi_{\bk \gamma} c_{\bk \gamma}$ is smooth and periodic (meaning that it depends on the gauge choice for $c_{\bk \gamma}$). This ensures that the BZ summation in the definition of $c_{\bR \gamma}$ is well-defined and that the states created by $c_{\bR \gamma}$ are exponentially localized (but they are generally not orthogonal; they correspond to coherent states studied by some of the authors in Ref.~\cite{Li2024constraints}). We also require $\chi_{\bk \gamma} c_{\bk \gamma}$ to reduce to $c^{\rm Hubbard}_{\bk \gamma}$ of the Hubbard model away from $\Gamma$, which describes the AA-localized part of the TBG Bloch states. The band topology, together with $C_{3z}$, then imply that $\chi$ must vanish as $\chi_{\bk \gamma} \propto k_x + i \gamma k_y$ near $\bk = \Gamma$ in any smooth gauge \cite{Li2024constraints}.  

We pause to comment that previous proposals for trions in TBG correspond to sharply distinct excitations. Indeed the former proposals correspond to generalized ferromagnetic spin polarons and transform under a higher representation of the flavor symmetry (e.g. $S=\frac{3}{2}$ for spin rotations)\cite{KhalafBabySkyrmions,schindlerTrionsTwistedBilayer2022}.
  
We will later derive the Green's function of $F^\dag_\bk$ and show that it precisely captures the missing state. The resulting spectral function is shown in Fig.~\ref{fig:blochsphere_trions}. We note that the trion is \emph{exactly} orthogonal to the flat band electron at $\Gamma$. 
This may be understood as a general consequence of band topology associated with the fact that the $F_\bk$ holes are, on their own, topologically trivial \footnote{The operators $F_\bk$ that we have defined are smooth and periodic in the BZ. This is not possible for creation operators of topological excitations.} We show this in two steps. First, the phase of the overlaps $\zeta_\bk^{\gamma,\gamma'} = \langle F_{\bk \gamma'} c^\dag_{\bk \gamma} \rangle$ wind around the BZ by $2\pi i \gamma$ in any smooth (non-periodic) gauge for $c^\dag_{\bk \gamma}$\cite{mishmashMajoranaLatticesQuantized2019,bultinckMechanismAnomalousHall2020}; each one must therefore vanish at some point in the BZ. Second, these points are pinned to $\Gamma$ by crystalline symmetries. Indeed, to be consistent with symmetry indicators of band topology \cite{FangGilbertBernevig, SymmetryIndicators}, $c_\Gamma$ must transform with a non-trivial representation relative to $F_\Gamma$. The latter point implies the excitations remain exactly orthogonal even when the Chern sectors are mixed by single-particle dispersion, including heterostrain if we assume spin-valley flavor symmetry. \footnote{TBG has a $\U(4)$ spin-valley rotation symmetry to an excellent approximation, even when single particle dispersion and strain are included \cite{BultinckHidden,VafekRG,TBGIVGroundState}. With this symmetry, we can define an intravalley $C_{2z}$ operator under which $F_{\Gamma, \gamma'}c^\dag_{\Gamma \gamma}$ is odd.} We emphasize that all of the previous statements are exact for any symmetric state and we therefore expect the trionic band edge to be rather robust to perturbations and deviations from the concentrated limit $s^2\ll 1$. We discuss how the trions imprint on the IKS state in the final section.

At charge neutrality, and in the limit of zero single-particle dispersion, the Green's function in the electron/trion space has a particularly simple form; it corresponds to an effective Dirac theory where the Dirac winding mixes electrons and trions
\begin{equation}
        \cG^{-1} = i\omega_n - \begin{pmatrix} 0 & \ov{\lambda}_{\bk \gamma} U \\ \lambda_{\bk \gamma} U & 0 \end{pmatrix}.
        \label{eq:cnpDiractheory}
    \end{equation}
    where $\lambda_{\bk \gamma}$ is the coefficient of the AA-localized part of the TBG Bloch state, which behaves as $\lambda_{\bk,\gamma} \propto k_x + i \gamma k_y$ for small $\bk$ around $\Gamma$ in any smooth gauge. Thus, the second term in \eqref{eq:cnpDiractheory} can be identified as an effective Dirac Hamiltonain in the electron-trion space.
    
The electron Green's function computed earlier projects onto the electron sector and only sees `half' of each Dirac cone, explaining the missing Dirac cone. More generally, $i\omega_n$ in \eqref{eq:cnpDiractheory} is replaced by a function $\cG_0(i\omega_n,\bk)$ which contains single particle and mean field type terms. These act to perturb and deform the electron-trion Dirac cone. As a simple example, in Fig. \ref{fig:blochsphere_trions}b, we included a small single-particle particle-hole breaking potential term that favors occupying the AA sites (see SI). Since the trions at $\Gamma$ are AA-localized while the $\Gamma$ electrons are not, this term acts as a Dirac mass term and makes the $\nu = 0$ valence band edge purely trionic. 

As we go away from neutrality,  the Hartree dispersion naturally acts in the same way: it favors occupying AA sites for $\nu < 0$ and de-occupying them for $\nu > 0$. As a result, the $\nu<0$ valence band edges are excited by trions. We plot the spectral functions for $\nu=-2$ in  Fig. \ref{fig:blochsphere_trions}c. For completeness, here we have also included a small BM dispersion associated with being slightly off the magic angle as well as an experimentally realistic level of heterostrain. However, these effects are largely confined to the electron-like conduction band; the trion band is remarkably insensitive to these single particle terms once it has been separated through the Hartree mass. 

We note that STS\cite{YazdaniCascade} and QTM\cite{shahalnewexpt} scans have observed an excitation band at a mostly fixed positive (negative) energy relative to the Fermi level for $\nu<0$ ($\nu > 0$) --- i.e. on the opposite side of doped charges. For electron-doping $\nu=-2$, we identify the band as the top most of Fig.\ref{fig:blochsphere_trions}c, with energy $2E_{\rm str}$. An analogous band will emerge continuously upon hole doping $\nu=-2$. Indeed, for excitations away from $\Gamma$, our strained model reduces to the effective Hamiltonian used to explain the feature in Ref. \cite{YazdaniCascade}. \\

\emph{\bf TBG Model and Energetic Hierarchy}---  Before computing the electron and trion spectra, we review TBG flat band model and hierarchy of scales discussed in Ref. \cite{ledwith2024nonlocal}. Our approach centers on two observations: (1) the gap separating the topological flat bands and the remote bands is significantly larger than the interaction scale \cite{xieSpectroscopicSignaturesManybody2019a, YazdaniCascade, Nuckolls_2023, EvaAndreiSTM2019, STMNadjPerge, BandFlattening,IlaniCascade,saitoIsospinPomeranchukEffect2021,bocarsly2024imagingcoulombinteractionsmigrating,shahalnewexpt} and (2) the charge density is concentrated at the AA sites \cite{xieSpectroscopicSignaturesManybody2019a, YazdaniCascade, Nuckolls_2023, EvaAndreiSTM2019, STMNadjPerge}. Due to (1), we study the interacting Hamiltonian projected to the flat bands \cite{BultinckHidden, ledwith2021strong, ledwith2024nonlocal}
\begin{equation}
  \H = \sum_\bk c^\dag_\bk h_\bk^{\rm sp} c_\bk + \H_{\rm int}, \quad \H_{\rm int} = \frac{1}{2A}\sum_\bq V_\bq \delta \rho_\bq \delta \rho_{-\bq},
  \label{eq:interactingham}
\end{equation}
Here, we have used a matrix notation where $c_{\bk}$ is a vector in the band label, $h^{\rm sp}_\bk$ is a matrix describing the single particle dispersion, and $\delta \rho_\bq = \rho_\bq - \ov{\rho}_\bq$ where $\rho_\bq = \sum_\bk c_\bk^\dagger \Lambda_{\bk,\bk+\bq} c_{\bk + \bq}$ is the projected density operator defined in terms of the form factor matrix $\Lambda^{\alpha \beta}_{\bk,\bk+\bq} = \langle u_{\alpha,\bk}|u_{\beta,\bk + \bq} \rangle$ and $\ov{\rho}_\bq = \frac{1}{2}\sum_{\bk,\bG} \delta_{\bq,\bG} \tr \Lambda_\bG(\bk) $ is half the total density of the flat bands. Throughout we use $\bk,\bk'$ for BZ momenta, $\bq,\bq'$ for unbounded momentum transfers, and $\bG,\bG'$ to denote reciprocal lattice vectors.

We will find it convenient to work in the so-called sublattice or Chern basis in which the two flat bands (per spin per valley) are decomposed into a $C = +1$ and a $C = -1$ band \cite{BultinckHidden, TBGIVGroundState}. In the Chern basis, the form factors have a ${\rm U}(4) \times {\rm U}(4)$ symmetry to good accuracy; deviations can be added perturbatively\cite{BultinckHidden,VafekRG,TBGIVGroundState}. In Ref.~\cite{ledwith2024nonlocal}, we derived an ansatz for the flat Chern band wavefunctions that faithfully accounts for their symmetry and topology and centers the concentration of the charge profile at the AA site, seen in STM measurements \cite{xieSpectroscopicSignaturesManybody2019a, YazdaniCascade, Nuckolls_2023, EvaAndreiSTM2019, STMNadjPerge} and consistent with the flat band wavefunctions close to the magic angle in the BM model \cite{BMpaper}. The wavefunctions in the $\gamma = +$ Chern sector are
\begin{equation}
  \psi_{\bk, \gamma=+}(\br) = \lambda_{\bk +} e^{i \bk \cdot \br} \sum_\bR \left(1 + \frac{(z-R)/\delta}{i k/s} \right)w_{\rm AA}(\br-\bR),
  \label{eq:fullWF}
\end{equation}
where $z=x+iy$, $k=k_x+i k_y$ is in the first BZ, and $w_{\rm AA}(\br) = \frac{e^{-\frac{\abs{\br}^2}{4\delta^2}}}{\sqrt{2\pi \delta^2}}\chi_0$ is a Gaussian at the $\br=0$ AA site of small width $\delta$. The wavefunctions in the other Chern sector $\gamma = -$ are related by $C_2 \T$: $\psi_{\bk -} = \overline{\psi_{\bk+}(-\br)}$. In the following, unless otherwise specified, we will be focusing on the $\gamma = +$ sector and defining $\lambda_\bk := \lambda_{\bk +}$. Results for the other sector are obtained by complex conjugation, $\lambda_{\bk -} = \overline{\lambda_\bk}$. Wavefunctions of the form \eqref{eq:fullWF} can be obtained from a variety of microscopic models \cite{LedwithFCI,LandauLevelWang,song2022TBGTHF} by taking the proper limit. 
They have been shown to match the BM wavefunctions for a fixed small value of $s$, which will be the small parameter of primary importance ($\delta$ mostly influences the overall interaction scale alongside the dielectric constant). It is convenient to choose units where the area of the unit cell and BZ are $2\pi$. For a fixed set of BM parameters, Ref. \cite{ledwith2024nonlocal} finds that $\delta=0.46, s=0.25$ matches the BM wavefunctions almost perfectly. Thus \eqref{eq:fullWF} opens the door to studying the interacting physics of TBG analytically, with $s^2 \ll 1$ as a small phase space parameter. 

We now discuss the prefactor and $\bk$-dependence of \eqref{eq:fullWF}. The prefactor satisfies $\abs{\lambda_{\bk}} = \frac{\abs{\bk}}{\sqrt{\abs{\bk}^2 + 2s^2}}$, such that \eqref{eq:fullWF} is normalized. The phase of $\lambda_{\bk}$ is a gauge choice. Because the band described by \eqref{eq:fullWF} is topological, there is no gauge that is globally smooth and periodic over the BZ. One particularly convenient gauge is
\begin{equation}
    \lambda_{\bk} = \abs{\lambda_{\bk}} = \frac{\abs{\bk}}{\sqrt{\abs{\bk}^2 + 2s^2}},
    \label{eq:singulargauge}
\end{equation}
which has a gauge singularity at $\Gamma$ from second term in \eqref{eq:fullWF}. We pause to comment that \eqref{eq:fullWF} with \eqref{eq:singulargauge} has discontinuities at the BZ boundaries suppressed by $s$ originating from expanding a function around $k=0$. They do not enter any observables to leading order in $s^2$, but it is convenient to remove them for numerics or plotting by replacing $k^{-1}$ with a periodic function that has a single pole at $k=0$. We choose the modified Weierstrass zeta function $\zeta(\bk)$ \cite{haldane2018modular}.
A smooth gauge in the vicinity of $\Gamma$ corresponds to $\lambda_{\bk} = \frac{k_x + i  k_y}{\sqrt{\abs{\bk}^2 + 2s^2}}$, which makes the winding manifest.

We pause to comment that $\lambda_{\bk \gamma}$ is a particularly natural choice for the function $\chi_{\bk \gamma}$ used to define the operators $c_{\bR \gamma}$ that enter the trion \eqref{eq:trioncreationtwo}. However, we will only demand that $\ov{\chi}_{\bk \gamma} \lambda_{\bk \gamma} \to 1$ for $\abs{\bk} \gg s$; this is sufficient to ensure that the AA sites in most of the band constructively interfere. 

An advantage of the gauge choice \eqref{eq:singulargauge} is that it makes manifest that \eqref{eq:fullWF} is equivalent to the Bloch wavefunction of a trivial AA-localized band for $\abs{\bk}\gg s$ not too close to $\Gamma$. Ref.~\cite{ledwith2024nonlocal} used this fact to construct Chern band Wannier states with almost all their weight in the central AA site; this led to parametrically small overlaps and exchange couplings despite the universal power law tail dictated by topology \cite{Li2024constraints}. More broadly, \eqref{eq:singulargauge} is convenient for the $s^2 \ll 1$ approximations which use the trivial part of the BZ away from $\Gamma$ as a momentum-space bath for the $O(s^2)$ $\Gamma$-electrons. We will use it exclusively below unless stated otherwise. Since in this gauge $\lambda_{\bk \gamma} = \ov{\lambda_{\bk \gamma}} = \lambda_\bk$, we can choose $\chi_{\bk \gamma}$ such that $\chi_{\bk \gamma} = \chi_\bk$. 

The contributions to $h^{\sp}$ as well as the form factors $\Lambda_{\bk,\bk+\bq}$ have explicit analytical forms derived in \cite{ledwith2024nonlocal}. For example, BM dispersion, heterostrain, and the particle-hole breaking dispersion enter as 
\begin{equation}
\begin{aligned}
    [h^{\rm BM}_{\bk}]_{\gamma \gamma'} & = E_{\rm BM}(1-\abs{\lambda_\bk}^2) \begin{pmatrix} 0 & e^{-2 i \theta_\bk} \\ e^{2 i \theta_\bk} & 0 \end{pmatrix}_{\gamma \gamma'}\\
    h^{\rm str}_\bk & = E_{\rm str}\abs{\lambda_\bk}^2 \gamma_x, \,\,\,\,h^{\rm PH}_\bk  = E_{\rm ph}(1-\abs{\lambda_\bk}^2),
    \end{aligned}
\end{equation}
where $e^{i \theta_\bk} = k/\abs{k}$ and the singularity of $h^{\rm BM}$ comes from the gauge choice. The full form factor is somewhat more complicated, but due to $s^2 \ll 1$ only contributions where at most one momentum $\bk$ is close to $\Gamma$ will enter:
\begin{equation}
    \Lambda^{\gamma \gamma'}_{\bk,\bk+\bq\to*} = \lambda_\bk \xi_\bq \begin{pmatrix}1 + \frac{\ov{q} \delta}{\ov{k}/s} & 0 \\ 0 & 1 + \frac{q \delta}{k/s}   \end{pmatrix}_{\gamma \gamma'}
\end{equation}
where $q = q_x+iq_y$, $\xi_\bq = e^{-\abs{q}^2 \delta^2/2}$ and we use $*$ to indicate a momentum that is taken to be away from $\Gamma$ (e.g. $\lambda_* =1$). Note that in the expressions above we have suppressed the spin and valley indices; all the matrices are the identity in these indices.\\

\begin{figure}
    \centering
    \includegraphics{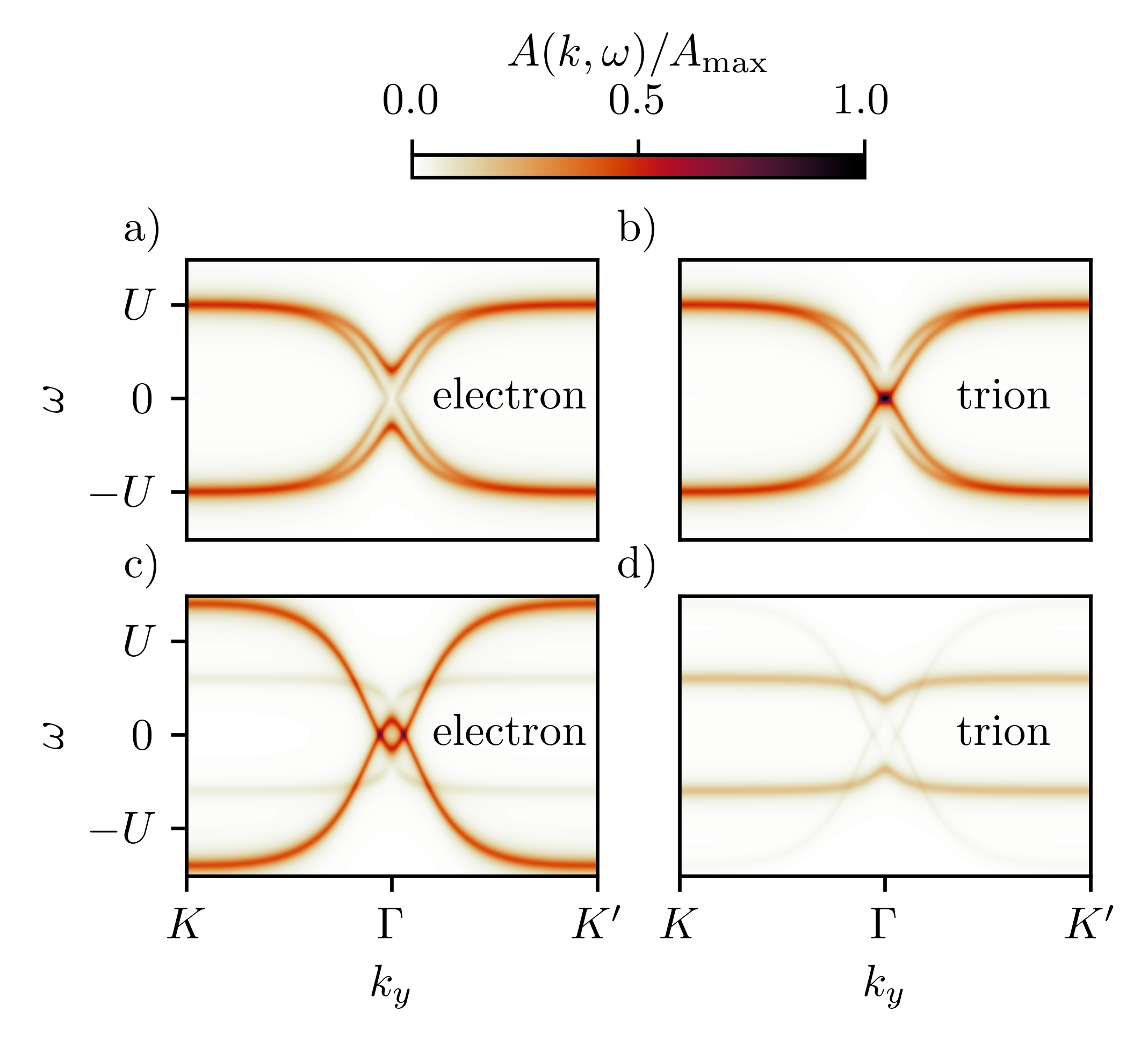}
    \caption{Spectral functions of electron and trion at charge neutrality: (a)-(b) show the spectral function for the Mott semimetal realized at temperatures above the ordering scale $U s^2$ and strain scale $E_{\rm str}$ (here we used $E_{\rm str} = 0$ and $E_{\rm BM} = 0.2 U$). (c)-(d) show the spectral function for the nematic semimetal realized at temperatures above $U s^2$ but below $E_{\rm str}$ ($E_{\rm str} = 0.3 U$, $E_{\rm BM} = 0.2 U$).}
    \label{fig:chargeneutrality}
\end{figure}

\emph{\bf Schwindger-Dyson equations for the Green's functions}--- Consider the Green's function $G^{cd}_{\tau,\bk} = -\langle \T c_{\bk,\tau} d_\bk^\dagger \rangle$ where $d_\bk$ is an arbitrary operator and $c_{\bk\tau} = e^{\tau \H} c_{\bk} e^{-\tau \H}$. We define the Fourier transformed Green's function via $G_{\bk,n} = \int d\tau e^{-i \tau \omega_n} G_{\bk,\tau}$. Taking the derivative in $\tau$, using the expression for the Hamiltonian (\ref{eq:interactingham}) and Fourier transforming to Matsubara frequencies yields (see SI for details)
\begin{equation}
    (i \omega_n - h_\bk^{\rm sp}) G^{cd}_{\bk,n} = \langle \{c_\bk, d_\bk^\dagger \} \rangle + G^{Od}_{\bk,n} 
    \label{SDcd}
\end{equation}
where $O_{\bk} = -[H_{\rm int}, c_\bk]$, given explicitly by
\begin{equation}
    O_{\bk} = \frac{1}{2A}\sum_{\bq} V_\bq \Lambda_{\bk,\bk+\bq}  \{\delta \rho_\bq, c_{\bk+\bq} \}
    \label{Ok}
\end{equation}
If we substitute $d_\bk = c_\bk$ in (\ref{SDcd}), we get the Green's function $G_{\bk,n} := G^{cc}_{\bk,n} = \frac{1}{i \omega_n - h_\bk^{\rm sp} - \Sigma_{\bk,n}}$ with the self energy $\Sigma_{\bk,n}$ given by
\begin{equation}
    \Sigma_{\bk,n} = G_{\bk,n}^{Oc} G^{-1}_{\bk,n} = G^{-1}_{\bk,n} G_{\bk,n}^{cO} 
    \label{SigmaGcO}
\end{equation}
where the second equality is obtained by applying the same steps to $G_{\bk}(\tau) = -\langle \T c_{\bk} c^\dag_{\bk,-\tau}\rangle$. This reduces the compuation of the electron Green's function to the computation of one of the Green's functions $G^{cO}$ or $G^{Oc}$.

For a more general operator $d$, we can split the Green's function $G^{Od}_{\bk,n}$ into a 1-particle irreducible (1PI) part $[G^{Od}_{\bk,n}]_{\rm 1PI}$ and a 1-particle reducible part (1PR) through $[G^{Od}_{\bk,n}]_{\rm 1PR} = G^{Oc}_{\bk,n} G^{-1}_{\bk,n} G^{cd}_{\bk,n} = \Sigma_{\bk,n} G^{cd}_{\bk,n}$, where we have used Eq.~\ref{SigmaGcO} in the last equality. Substituting in (\ref{SDcd}) leads to the relation
\begin{equation}
    G^{-1}_{\bk,n} G^{cd}_{\bk,n} = \langle \{c_\bk, d_\bk^\dagger \} \rangle + [G^{Od}_{\bk,n}]_{\rm 1PI}
    \label{SD1PI}
\end{equation}
This is one of the fundamental relations that we will use in our derivation. Since the self-energy $\Sigma_{\bk,n}$ is directly related to $G^{\rm cO}$, we see that substituting $d_\bk = O_\bk$ in Eq.~\ref{SD1PI} and using (\ref{SigmaGcO}) yields $\Sigma_{\bk,n} = \langle \{c_\bk, O_\bk^\dagger \} \rangle + [G_{\bk,n}^{OO}]_{\rm 1PI}$. The first term can be simplified by noting that $\langle \{c_\bk, O^\dagger_\bk \} \rangle = h_\bk^{\rm mf}(Q) = h^H_\bk(Q) + h^F_\bk(Q)$, where $Q_{\alpha \beta}(\bk) = \langle [c_{\beta \bk}, c^\dag_{\alpha \bk}]\rangle$ is the order parameter and the Hartree and Fock Hamiltonians are given by 
\begin{equation}
\begin{aligned}
  h^H_\bk(Q) & = \frac{1}{A}\sum_\bG V_\bG \Lambda_{\bk,\bk + \bG} \sum_{\bk'} (\bk') \tr \Lambda_{\bk',\bk' - \bG} Q_{\bk'}\\
  h^F_\bk(Q) & = \frac{1}{2A} \sum_\bq V_\bq \Lambda_{\bk,\bk-\bq} Q_{\bk-\bq} \Lambda_{\bk-\bq,\bk},
  \end{aligned}
  \label{eq:meanfieldterm}
\end{equation}
This leads to the expression for the self-energy
\begin{equation}
    \Sigma_{\bk,n} = h^{\rm mf}_\bk(Q) + [G^{OO}_{\bk,n}]_{\rm 1PI}
\end{equation}

Up to this step, all our manipulations have been exact and we have not used the small phase space approximation $s^2 \ll 1$. The main simplification of this approximation is that it allows us to assume that every momentum that is summed over is away from $\Gamma$, denoted by $``*"$. As a result, any 1PI correlation function can be evaluated in terms of a 1PI in a trivial classical model defined by the replacements $\Lambda_{\bk,\bk+\bq} \to \xi_\bq$, $h^{\rm sp}_\bk \to h^{\rm sp}_*$: \begin{equation}
  \H_{\rm cl} = \sum_\bR c^\dag_\bR h^\sp_* c_\bR +  \sum_{\bR,\bR'} U(\bR-\bR') \delta n_\bR \delta n_{\bR'},
  \label{eq:Hclassical}
\end{equation}
where $U(\bR) = \frac{1}{2A}\sum_\bq V_\bq \abs{\xi_\bq}^2 e^{i \bq \cdot\bR}$ and $\delta n_\bR = n_\bR - 4$. All the non-trivial momentum dependence is encoded in the external legs of the 1PI diagram. We note that the limits $T\to 0$ and $s^2 \to 0$ do not commute; for example ferromagnetic exchange interactions $J \sim Us^2$ can lead to ordering at $T\ll Us^2$ whereas $T \gg Us^2$ must be disordered \cite{ledwith2024nonlocal}. We therefore only trust these results in the Mott regime $U \gg T \gg Us^2$. \\

For the operator $O_{\bk,n}$, we note that summation on $\bq$ in Eq.~\ref{Ok} means that all fermion operators can be taken away from $\Gamma$. This means that the density operators simplify to $\delta \rho_\bq = \frac{\xi_\bq}{\sqrt{N}}\sum_\bR e^{-i \bq \cdot \bR} \delta n_\bR$. The expression for $[G^{OO}_{\bk,n}]_{\rm 1PI}$ then becomes 
\begin{equation}
    [G^{OO}_{\bk,n}]_{\rm 1PI} = U^2 |\lambda_\bk|^2 \Xi_{n}, \quad  \Xi_{n} = \langle \{ c_\bR, \delta n_\bR\} \{\delta n_{\bR}, c^\dag_\bR\} \rangle_{\rm 1PI}
\end{equation}
where $\Xi_{n}$ is evaluated in the classical model defined by the Hamiltonian (\ref{eq:Hclassical}) and we used the relation
 $\frac{1}{A}\sum_{\bq}V_{\bq}\Lambda_{\bk,\bk+\bq}\xi_{-\bq'} = U \lambda_\bk$ which assumes the interaction only depends on $|\bq|$.
 The classical correlation function can in principle be computed at any filling, but it is particularly simple at integer fillings where the thermal density matrix is an eigenstate of $\delta n_\bR$ with eigenvalue $\nu$, which is our main focus here. This yields the expression
\begin{equation}
   [G^{OO}_{\bk,n}]_{\rm 1PI} = U^2 |\lambda_\bk|^2  \Xi_{n}, \quad  \Xi_{n} = \frac{1-Q_*^2}{i\omega_n + \mu_\nu - Q_*U}
\end{equation}
where $\mu_\nu = \mu - 2\nu U$ is the chemical potential relative to the Hartree shift at filling $\nu$. The resulting electron's Green function is
\begin{equation}
  G_{\bk,n} = \frac{1}{[G^{\rm mf}_{\bk,n}]^{-1} - U^2 |\lambda_\bk|^2 \Xi_{n}}.
  \label{eq:electronGFgeneral}
\end{equation}
where $[G^{\rm mf}_{\bk,n}]^{-1} = i \omega_n + \mu - h^{\rm sp}_\bk - h^{\rm mf}_\bk(Q)$. This formula is very general; it can describe the symmetry-broken state as well as the thermal state at different fillings by choosing appropriate $Q_*$.  The poles of $G_{\bk,n}$ are obtained by solving the equation $U^2 |\lambda_\bk|^2 \Xi_n G_{\bk,n}^{\rm mf} = 1$. If we focus close to the $\Gamma$ point where $\lambda_\bk \sim \bk^2$, we find that the poles of $G_{\bk,n}$ come either from poles of $G^{\rm mf}_{\bk,n}$ or $\Xi_n$. For a pole $i \omega_n = z^{\rm mf}$ of $G^{\rm mf}$, we can write $G_{\bk,n}^{-1} = i \omega_n - z^{\rm mf} + O(\bk^2)$, which gives a pole of $G_{\bk,n}$ at $i \omega_n = z^{\rm mf} + O(\bk^2)$ with residue $Z^{\rm mf} = 1$. On the other hand, close to a pole $i \omega_n = z^{\Xi} + O(\bk^2)$ of $\Xi_n$, we have $G_{\bk,n} \approx (i \omega_n - z^\Xi)/((i \omega_n - z^\Xi) \alpha - U^2 \bk^2/(2s^2))$, for some constant $\alpha$, which gives a pole at $i \omega_n = z^\Xi + O(\bk^2)$ with residue $Z^\Xi_\bk \sim \bk^2$ which vanishes as $\bk \rightarrow 0$.

It is instructive to comment on the properties of $G_{\bk,n}$ at charge neutrality $\nu = 0 = \mu$, where its form significantly simplifies. In the ordered state (ferromagnet or nematic semimetal), $Q_*^2 = 1$ leading to $\Xi = 0$ which implies $G_{\bk,n} = G^{\rm mf}_{\bk,n}$ as expected. In the thermally disordered state, $Q_* = 0$, $h_\bk^{\rm mf} = 0$, and $\Xi_n = 1/\omega_n$ leading to
\begin{equation}
  G_{\bk,n} = \frac{1}{i \omega_n - h_\bk^{\rm sp} - |\lambda_\bk|^2 \frac{U^2}{i\omega_n}}.
  \label{eq:neutrality}
\end{equation}
This has poles at $i \omega_n = z^\pm_\bk = \frac{1}{2}(h_\bk^{\rm sp} \pm \sqrt{[h_\bk^{\rm sp}]^2 + 4U^2 |\lambda_\bk|^2})$. Away from $\Gamma$, we can replace $|\lambda_\bk| \rightarrow |\lambda_*| = 1$, and get two Hubbard bands at $\pm U$ when $h_*^{\rm sp} = 0$, which can be further split at finite $h^{\rm sp}_*$. For $\bk \approx 0$, we have $|\lambda_\bk| \approx \bk/(\sqrt{2}s)$ while $h_\Gamma^{\rm sp}$ is generally non-zero. This leads to two poles of different character; $\omega_n = z^+_\Gamma \approx h_\Gamma^{\rm sp}$, yields a band whose gap is controlled by the single particle dispersion at $\Gamma$ with quasiparticle residue $Z^+_\bk$ approaching $1$ at $\bk \rightarrow 0$, whereas $\omega_n = z^-_{\bk \approx 0} \sim \bk^2$ describes a quadratic band touching with quasiparticle residue $Z^-_\bk \rightarrow 0$ as $\bk \rightarrow 0$. The spectral function for the thermal state at charge neutrality is shown in Fig.~\ref{fig:chargeneutrality} confirming this analysis.\\

\emph{\bf Trion Green's function}--- While the main goal of the calculation above was to obtain the electron Green's function, an interesting by-product is the calculation for the Green's function of the operator $O_\bk$, which can be written by including the 1-particle reducible piece
\begin{equation}
  G^{OO}_{\bk,n} = \Sigma_{\bk,n} (1 + G_{\bk,n} \Sigma_{\bk,n}) - h^{\rm mf}_\bk  
  \label{eq:motivatingOOdag}
\end{equation}
For the thermal state at $\nu = 0$, we get
\begin{equation}
  G^{OO}_{\bk,n} = |\lambda_\bk|^2  U^2 \frac{1}{i\omega_n - |\lambda_\bk|^2  \frac{U^2}{i \omega_n-h_{\rm sp} }}.
  \label{GOO}
\end{equation}
We see that, compared to (\ref{eq:neutrality}), this has the same poles at $z_\bk^\pm$, but their quasiparticle residues has swapped such that $Z_\bk^+ \rightarrow 0$ and $Z_\bk^- \rightarrow 1$ as $\bk \rightarrow 0$. This seems to suggest that the operator $O_{\bk,n}$ creates the particles that carry the missing spectral weight. There are, however, two issues in interpreting this as the Green's function for the missing excitation. First, the overall factor $W_\bk$ means that this object identically vanishes at $\bk = 0$ at any frequency. Second, the operator $O_\bk$ has complicated anticommutation relationship with $c_\bk$ making it difficult to interpret it as a distinct excitation. This motivates the definition of the trion operator given in Eq.~\ref{eq:trioncreationtwo}. We see that this operator differs from $O_{\bk,n}$ in two ways: (i) we have removed the factor $V_\bq \Lambda_{\bk,\bk+\bq}$ which is responsible for the factor $U^2 |\lambda_\bk|^2$ in (\ref{GOO}) and (ii) we have introduced an extra single particle term with $A_\bk = \chi_\bk(2\langle \delta n_\bR\rangle - \frac{1}{N}\sum_{\bk'} \abs{\chi_\bk}^2 Q_{\bk+\bk'})$ defined so that $\langle \{ c_\bk, F^\dag_\bk \} \rangle = 0$ (see SI). Small $s^2 \ll 1$ implies $A_\bk \approx (2\nu - Q_*)$. 

As before, we can split the trion Green's function into its 1PI and 1PR parts; $G^{FF}_{\bk n} = [G^{FF}_{\bk n}]_{\rm 1PI} + G^{Fc}_{\bk n} G^{-1}_{\bk,n} G^{cF}_{\bk,n}$. The 1PI part $[G^{FF}_{\bk n}]_{\rm 1PI} = \Xi_n$ is classical. For 1PR part, we use Eq.~\ref{SD1PI} with $d_\bk = F_\bk$ which yields 
\begin{equation}
  G_{\bk,n}^{-1} G^{cF}_{\bk,n} = [G^{OF}_{\bk,n}]_{\rm 1PI} = U \lambda_\bk \Xi_n
  \label{eq:GcF}
\end{equation}
The above equation, and its Hermitian conjugate, can be inserted into the 1PR part of $G^{FF}_{\bk,n}$ to obtain
\begin{equation}
  G^{FF}_{\bk,n} = \Xi_n + U^2 \abs{\lambda_\bk}^2 \Xi_n G_{\bk,n} \Xi_n 
   = \frac{1}{ \Xi_n^{-1} - U^2 \abs{\lambda_\bk}^2 G^{\rm mf}_{\bk,n} }.
  \label{eq:trionGF}
\end{equation}
We see that this has the same poles as Eq.~\ref{eq:electronGFgeneral} at $U^2 |\lambda_\bk|^2 G_{\bk,n}^{\rm mf} \Xi_n = 1$ but with switched resides, such that $Z_\bk \to 0$ pole for the electron Green's function is now $Z_\bk = 1$ for the trion Green's function and vice versa. Thus, we can identify the missing weight in the electron spectral function with trion excitations that are exactly orthogonal to the electron at the $\Gamma$ point.

To isolate the self energy we compute the Green's function inverse
\begin{equation}
  \cG_{\bk,n}^{-1} = \begin{pmatrix} G^{cc} & G^{cF} \\ G^{Fc} & G^{FF} \end{pmatrix}^{-1} = \begin{pmatrix} [G^{\rm mf}_{\bk,n}]^{-1} & -\lambda_\bk U \\ -\ov{\lambda}_\bk U & \Xi_n^{-1} \end{pmatrix}.
  \label{eq:DiracinverseG}
\end{equation}
which is straightforward to verify. This equation is valid in any gauge, but it's useful to consider a smooth but non-periodic gauge where $\lambda_\bk \approx k_x + i k_y$ at small $\bk$. Then, we see that the electrons and trions interact through an interaction-generated Dirac self energy. Mean-field and single particle terms can act as Dirac masses. In the mean-field limit $Q_*^2 \to 1$ we have $\Xi \to 0$ such that the total density of states of trions vanishes and $G_{cc} = G_{\rm mf}$ as expected.

The electron and trion spectral functions in the Mott regime $Us^2 \ll T \ll U$ are shown in Figs.~\ref{fig:blochsphere_trions} and \ref{fig:chargeneutrality}. The simplest limiting case to understand is charge neutrality in the absence of single-particle dispersion or strain. In this case, the electron and trion form a Dirac cone where the spinor winds in the electron-trion plane such that half of the spectral weight goes to each. It is important to note here that the electron operator already has a built-in phase winding due to the topology of the parent band while the trion operator has no phase winding because it is built of local topologically trivial (away from $\Gamma$) operators. As a result, there is no net Berry phase away from $\Gamma$ due to the cancellation between the electron winding and the Dirac winding. This is an important consistency check since the operators away from $\Gamma$ look like trivial Bloch waves of localized states. Adding small particle-hole breaking, which favors AA occupation, acts as a Dirac mass which makes the spinor point up (the electron) in the conduction band at $\Gamma$ and down (the trion) in the valence band, rotating gradually towards the equator as we move away from $\Gamma$. This explains the vanishing electron spectral weight observed earlier and makes it clear that the trion exactly compensates for it.

In Fig.~\ref{fig:chargeneutrality}, we show the electron and trion spectral weights at charge neutrality at two different temperatures. Panels (a) and (b) show the spectral functions for temperatures above the strain scale $U s^2, E_{\rm str} \ll T \ll U$. This corresponds to the Mott semimetal state where the nonlocal moments are disordered. Compared to Fig.~\ref{fig:chargeneutrality}, we include a single-particle dispersion $E_{\rm BM} = 0.2 U$. This splits the electron-trion Dirac cone into a gapped cone containing all the electron spectral weight at $\Gamma$ and a gapless quadratic band touching containing all the trion spectral weight at $\Gamma$. In panels (c) and (d), we consider temperatures below the strain scale: $T = 0.6 E_{\rm str}$, $E_{\rm str} = 0.3 U$. The spectral function describes a strain-stabilized nematic semimetal  \cite{liuNematicTopologicalSemimetal2021,parkerStrainInducedQuantumPhase2021}. The small trion spectral weight arises from the thermal fluctuations and would vanish at zero temperature.

In Fig.~\ref{fig:blochsphere_trions}c, we show the spectral functions for the electron and trion at $\nu = -2$ for temperatures $U s^2 \ll T \ll E_{\rm str}, U$ and $E_{\rm str}=0.3U$, $E_{\rm BM} = 0.2U$. We see that the missing electron spectral weight at the valence band edge at $\Gamma$ is perfectly compensated by an enhanced spectral weight in the trion spectral function. Away from $\Gamma$, the electron conduction band is split due to strain. The total weight in the electron conduction and valence band is 6 and 2 respectively, up to order $s^2$ corrections. This is consistent with filling 2 out of 8 bands. The total weight in the trion conduction and valence band is 2 and 2, respectively. The reduced trion weight corresponds to a vanishing spectral weight in the mean-field-like conduction band selected by strain. The strain-selected valence band is similar to the case of charge neutrality with the Hartree dispersion naturally playing the role of the particle hole breaking dispersion. Importantly, we see that there is a reduction in the trion density of states as we move away from neutrality, such that we are effectively considering $4 - |\nu|$ electron-trion Dirac cones around filling $\nu$.\\

\emph{\bf Physical origin of the trions}--- We would like to discuss the physical origin and some of the remarkable properties of the trion excitation. First, we note that the trion is a light particle. This is remarkable given that the trion consists entirely of heavy objects with infinite effective mass. 

One way to understand why the trion forms and why its mass is light is the following. As discussed in Ref.~\cite{ledwith2024nonlocal}, the Wannier functions of the model feature a central peak, containing most of the weight of the wavefunction, and a power-law tail with a parametrically small weight $\sim s^2$. The interaction predominantly consists of a Hubbard part which comes from the main peaks. However, important effects of the topology are encoded by taking the long-range parts of the interaction arising from the tails. The leading order term is the one with three peaks and a tail corresponding to three particles at the same site and a particle at a far away site. Such term effectively acts as a long-range electron-trion hopping term that turns a trion at $R$ into an electron at $R'$ at distance $\sim 1/s$ from $R$ with hopping amplitude $s^2$. Integrating out the electron yields a long-range trion hopping process of amplitude $U s^2$ and range $1/s$. The electron-trion hopping function is essentially the Fourier transform of the function $U \lambda_\bk$ which gives a trion dispersion $\sim U |\lambda_\bk|^2$ with minimum at $\Gamma$ and bandwidth $\sim U$. Thus, the trion binding is kinetic in origin and arises from their ability to delocalize over a large area $\sim 1/s^2$ through the small but long-ranged part of the interaction. It also implies the trion is light with an effective mass of the order $\sim s^2/U$ around the $\Gamma$ point that gets quickly heavier as we move away from $\Gamma$. The origin of the light mass can be traced back to the concentrated quantum geometry and topology of the model that gives rise to the long-range part of the interaction and ensures the strength of the long-range electron-trion hopping and its effective range are related. \\

\emph{\bf Intervalley Kekul\'e Spiral}--- Thus far we have shown that the charged excitations of the TBG normal state $(\nu \lesssim -2, T\gtrsim Us^2)$ are Dirac trions. It is pertinent to study their fate as $T \to 0$, especially for $\nu < -2$ where the superconductor forms. The $T=0$ ground states, particularly around $\nu = -2$, have been extensively studied. Two types of ground states have been proposed. In strain-free TBG near the magic angle, the ground state is a generalized quantum Hall ferromagnet corresponding to filling two out of eight bands in the Chern basis \cite{BultinckHidden, TBGIVGroundState}. These are translationally symmetric mean field states with no missing electron state at $\Gamma$. Furthermore, they are annihilated by the trion operator $F^\dag_\bk$ discussed here (they may admit energetically favorable spin polaron or skyrmion excitations on doping the \emph{conduction} band \cite{KhalafBabySkyrmions, KwanSkyrmion}). In this section, we will instead focus on the IKS state which was proposed \cite{KwanKekule} and observed \cite{Nuckolls_2023} as the ground state of strained TBG at integer $\nu \neq 0$. 

We begin by deriving the mean-field IKS state analytically through $s \ll 1$. We then compute its momentum resolved spectral function $A(\bk,\omega) = \frac{1}{\pi} \rm{Im}\langle \T c_{\bk n}c^\dag_{\bk n} \rangle$, relevant for ongoing momentum resolved tunneling experiments \cite{QTM}. We pause to comment on a subtle aspect of this quantity. For translationally-symmetric mean-field states, the spectral function is $\sum_n\delta(\omega - E_{n}(\bk))$ where $E_n(\bk)$ are the Hartree Fock energy bands. This is no longer the case if the state breaks translations since $\bk$ no longer labels the energy eigenstates. Instead, the eigenoperators are of the form $u_\bk c^\dag_\bk + v_\bk c^\dag_{\bk + \bq_s}$, where $\bq_s$ is a translation breaking wavevector. Then $A(\bk,\omega) \propto \abs{u_\bk}^2 = Z_\bk$ can correspond to a quasiparticle residue different from unity.  

We find that the IKS state has a $Z_\bk \to 0$ state at the $\Gamma$-point valence band edge, just like the thermal state in Fig. \ref{fig:blochsphere_trions}. As before, the trion operator $F^\dag_\Gamma$ is orthogonal to any $\Gamma$-point electron operator and it precisely creates this state. 
Due to the translation breaking, however, $F^\dag_\bk$ could act equivalently to an electron operator at a \emph{different} momentum $c^\dag_{\bk \pm \bq_s}$. 
Indeed, this turns out to be the case; the missing $\Gamma$ point state can also be created by $c^\dag_{\bk \pm \bq_s}$ where $q_s$ is the IKS wavevector. At a high level, we can understand the compatibility of the two pictures as follows. The trion is composed of an electron and an electron-hole pair at total momentum $\bk$. Then, in a translation-breaking state, an electron hole pair with momentum $\pm q_s$ can act as a $c$-number order parameter. This results in the trion $F^\dag_\bk$ acting like the remaining electron $c^\dag_{\bk \mp \bq_s}$.\footnote{Note that here the electron-hole pair is not that of the density; instead one should write $c^\dag_{\bR \alpha} n_\bR = c^\dag_{\bR \alpha} \sum_\beta c^\dag_{\bR \beta} c_{\bR \beta}$ and contract $\langle c^\dag_{\bR \alpha} c_{\bR \beta} \rangle$ to obtain the IKS order parameter.}

\begin{figure}
    \centering
    \includegraphics{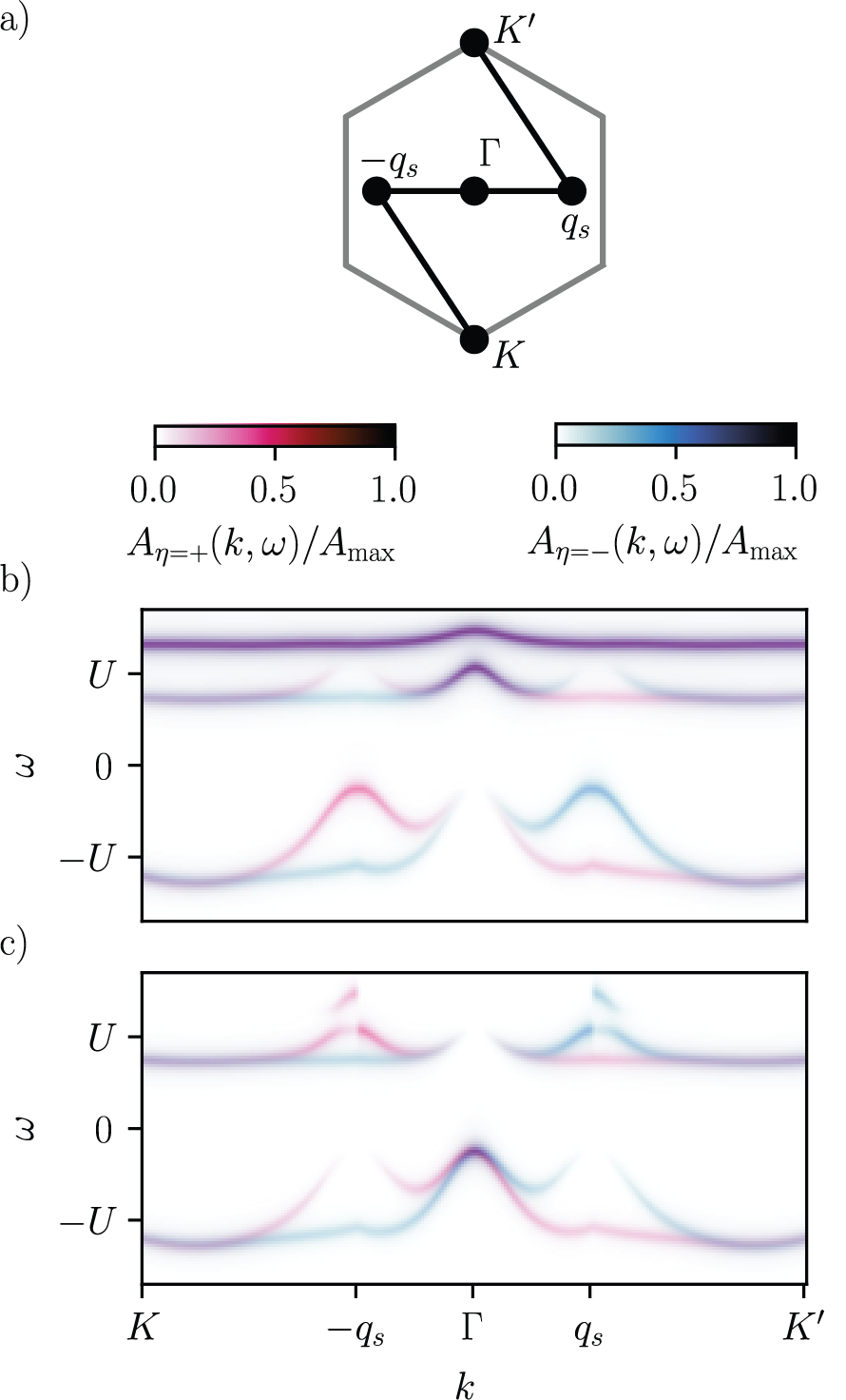}
    \caption{(b) Electron and (c) Trion spectral functions, resolved by valley (red vs blue color scales) in the intervalley Kekule spiral (IKS) state with wavevector $\bq_s$ plotted along momentum cut shown in (a). Here, we use $\bq_s = 0.38 \bG_x$ for definiteness, where $\bG_x$ is the reciprocal lattice vector in the $x$-direction. We also used $E_{\rm BM} = 0.2 U$, $E_{\rm str} = 0.3 U$.}
    \label{fig:IKS}
\end{figure}

We now derive the IKS state explicitly and verify our claims above. We begin by observing that low-energy mean-field states in TBG have order parameters $Q_{\bk,\bk'}$ which, for $\bk,\bk'$ away from $\Gamma$, correspond to mean-field ground states of the trivial classical model \eqref{eq:Hclassical}. Indeed, to leading order in $s^2$, contributions to the Hartree Fock energy predominantly come from momenta away from the $\Gamma$ point. The mean-field ground states of \eqref{eq:Hclassical} have definite and equal occupation of each lattice site, which translates to $Q_{\bk,\bk'} = Q(\bk-\bk')$ and $\tr Q(\bk-\bk') \propto \delta_{\bk,\bk'}$. The IKS state forms in the valence band of strained TBG, here $\gamma_x = -1$. It has a spiral order in the phase associated with intervalley coherence with a single wavevector $\bq_s$ with no net valley polarization or spin order.  The IVC spiral corresponds to boosting one valley relative to the other with the momentum $\bq_s$. These facts all taken together fix, up to an overall $U(1)$ valley rotation,
\begin{equation}
\begin{aligned}
    Q_{*}(\bk-\bk') & = -P_{\gamma_x=+}\delta_{\bk,\bk'} + Q_{s *}(\bk-\bk')\\
    Q_{s *}(\bk-\bk') & = P_{\gamma_x=-}\begin{pmatrix} 0 & -\delta_{\bk+\bq_s,\bk'} \\-\delta_{\bk,\bk'+\bq_s} & 0 \end{pmatrix}_{\rm valley}.
    \label{eq:IKSorderparam}
    \end{aligned}
\end{equation}
where we included a $*$ in the subscript of $Q$ to indicate that the expression should be used for momenta away from $\Gamma$. The wavevector $\bq_s$ is only very weakly pinned and it can take on a variety of nonzero values; for the spectral functions in Fig.~\ref{fig:IKS}, we pick $\v q_s = 0.38 \v G_x$, where $\v G_x$ is the reciprocal lattice vector in the $+k_x$ direction, but our conclusions are independent of this choice as long as $\v q_s$ is not too small.

The first term in $Q$ is equivalent to $Q_{\rm th}$, the HF correlator associated with the $\nu = -2$ thermal state. It leads to a corresponding Hartree Fock dispersion $h_{\rm mf, th}$. Let us define $G_{\rm mf, th}^{-1}(\bk) = i \omega_n +\mu - h_{\rm sp}(\bk) - h_{\rm mf, th}(\bk)$, which is equal to $G_{\rm mf}$ in the thermal $\nu=-2$ state. 

We now calculate the mean-field Hamiltonian associated to the spiral part of the IKS order parameter. It has the form 
\begin{equation}
    H_s = \sum_\bk \Psi^\dag_\bk P_{\gamma_x=-}\begin{pmatrix} 0 & h_s^\dag(\bk) \\ h_s(\bk) & 0\end{pmatrix} \Psi_\bk
\end{equation}
where $h_s^\dag(\bk) = \frac{1}{2A}\sum_\bq V_\bq \Lambda_{\bk,\bk+\bq} \Lambda_{\bk+\bq+\bq_s,\bk+\bq_s}$ and $\Psi = \begin{pmatrix} c_{\bk \eta=+} &  c_{\bk + \bq_s,\eta=-} \end{pmatrix}^T$. For $\bk+\bq_s$ away from $\Gamma$, but $\bk$ allowed near $\Gamma$, we can replace the second form factor in $h_s$ with $\xi_\bq$ such that $h_s^\dag(\bk) = h_s(\bk) =  \lambda_\bk U$.

The full IKS Green's function $\mathcal{G}_{\rm IKS} = -\langle \T \Psi \Psi^\dag \rangle$ can then be written in the $\Psi$ basis as
\begin{equation}
    \mathcal{G}_{\rm IKS}^{-1}(\bk) = \begin{pmatrix} G_{\rm mf, th}^{-1}(\bk) & - P_{\gamma_x=-1}\lambda_\bk U \\ -P_{\gamma_x=-1}\lambda_\bk U & G^{-1}_{\rm mf, th}(*)\end{pmatrix},
    \label{eq:IKSasTrion}
\end{equation}
for $\bk + \bq_s$ away from $\Gamma$. For $\bk' = \bk+\bq_s$ near $\Gamma$ (and $\bk$ far from $\Gamma$) one can evaluate $\mathcal{G}_{\rm IKS}^{-1}(\bk)$ analogously. 
It is useful for numerics and plotting to write a general expression that reduces to two cases above upon taking $\bk + \bq_s \to *$ and $\bk \to *$ respectively. This can be obtained by replacing $G_{\rm mf, th}(*) \to G_{\rm mf, th}(\bk + \bq_s)$ and $\lambda_\bk U \to \lambda_\bk \lambda_{\bk + \bq_s} U$ in \eqref{eq:IKSasTrion}.

We emphasize that \eqref{eq:IKSasTrion} and \eqref{eq:DiracinverseG} have an essentially equivalent form. Indeed, for the $\nu = -2$ thermal state with $E_{\rm str} \gg T$ the trion block, $\Xi^{-1}$, in \eqref{eq:DiracinverseG} is infinite in the $\gamma_x = +1$ sector, corresponding to vanishing trion spectral weight. Thus the off-diagonal couplings $\lambda_\bk U$ can be replaced by those in \eqref{eq:IKSasTrion}. Furthermore, $\Xi = G_{\rm mf, th}(*)$ in the $\gamma_x = -1$ sector of interest. Thus, for $\bk+\bq_s$ away from $\Gamma$, \eqref{eq:IKSasTrion} precisely maps onto \eqref{eq:DiracinverseG} under the identification $c_{\bk+\bq_s,\eta=-} \leftrightarrow F_{\rm th,\bk, \eta=+}$, where $F_{\rm th,\bk, \alpha} = \frac{1}{\sqrt{N}}\sum_\bR e^{-i \bk \cdot \bR} \{c_{\bR, \alpha}, \delta^\nu n_\bR \} + [Q_{\rm th} c_\bk]_\alpha $ is the trion operator for the thermal state.

The above discussion suggests that the thermal trion operator acts on the IKS state as an electron operator at a momentum shifted by $\pm q_s$. We can verify this explicitly through showing that the trion operator $\tilde{F}$ has zero spectral weight. We define
\begin{equation}
\begin{aligned}
    \tilde{F}_{\bk \eta} & = F_{\rm th,\bk \eta} + \sum_{\bk'}[Q_{s*,\bk-\bk'} c_{\bk'}]_\eta\\
     &= F_{\rm th,\bk,\eta} - c_{\bk+\eta\bq_s, -\eta} 
\end{aligned}
\label{tildeF}
\end{equation}
such that $\langle\{\tilde{F}_\bk, c_{\bk'} \} \rangle_{\rm IKS} = 0$.
We only specified the valley label since all quantities of interest are in the $\gamma_x = -1$ valence bands and nothing depends on spin. We then calculate
$G^{\tilde{F} \tilde{F}} = [G^{\tilde{F} \tilde{F}}]_{\rm 1PI} - G^{\tilde{F} c} [G^{cc}]^{-1} G^{c \tilde{F}}$, where now the matrix multiplications include momentum indices $\bk,\bk'$ and corresponding sums as well because the Green's functions are no longer diagonal in momentum. Analogously to the thermal trion derivation, we obtain $[G^{\tilde{F} \tilde{F}}_{\bk,\bk'}]_{\rm 1PI} = \Xi(i\omega_n) \delta_{\bk,\bk'}$ and $\left([G^{cc}]^{-1} G^{c \tilde{F}}\right)_{\bk,\bk'} = [G^{Oc}_{\bk,\bk'}]_{\rm 1PI} = \lambda_\bk \Xi(i\omega_n)$, where as before $\lambda_\bk$ comes from the form factor in $O_\bk$. Since the IKS state is a Slater determinant, with $Q_*^2 = 1$, we have $\Xi(i\omega_n) \propto 1-Q_*^2 = 0$ such that $G^{\tilde{F} \tilde{F}} = 0$. This implies that $\tilde{F}$ annihilates the IKS ground state.

We therefore conclude that the trion of the thermal state becomes a momentum-shifted electron operator upon the onset of IKS order: $F_{\rm th,\bk\eta}\ket{\Psi_{\rm IKS}} = c_{\bk+\eta\bq_s, -\eta} \ket{\Psi_{\rm IKS}}$. We can see this explicitly in Fig. \ref{fig:IKS}b,c. In panel b, we see that the electron spectral function has a missing state at $\Gamma$ that can be excited with the operator $F_{\rm th, \bk}$ (panel c), just like the thermal state. In contrast to the thermal state the operators $c_{\bk - \eta \bq_s, \eta}$ also suffice, as indicated by the bright band edges at $\pm q_s$ in valleys $\eta=\mp$. We emphasize that in both the thermal and IKS states the excitations are light carriers that come from the AA-localized part of the band.\\

\emph{\bf Discussion and conclusion}--- We begin by comparing and contrasting the Dirac trions presented here to other composite excitations in strongly correlated systems. Thus far in TBG, the composite excitations most studied are those related to charged skyrmions on top of the generalized ferromagnetic strong coupling insulators \cite{SondhiSkyrmions,Chatterjee19,SkPaper,KwanSkyrmion}. These correspond to electrons bound to $N$ (generalized) spin flips for any integer $N \geq 1$. For $N=1$, one obtains a ferromagnetic spin polaron excitation consisting of an electron or hole plus a single (generalized) spin flip \cite{KhalafBabySkyrmions,schindlerTrionsTwistedBilayer2022,prichardDirectlyImagingSpin2024,lebratObservationNagaokaPolarons2024,whiteDensityMatrixRenormalization2001,maskaEffectiveApproachNagaoka2012,grusdtMicroscopicSpinonchargonTheory2019,davydovaItinerantSpinPolaron2023}. This can also be viewed as a trion \cite{KhalafBabySkyrmions,schindlerTrionsTwistedBilayer2022} since a spin flip can be viewed as a hole in the majority spin band plus an electron in the minority spin band.
Such excitations are sharply distinct from the trion we discuss here due to their (generalized) spin; e.g. for a $z$-polarized ferromagnet the trion has spin $S_z = 3/2$. They are also typically heavier than the electron (though pairs of them can be light \cite{SkPaper, KwanSkyrmion}). Trions were also used in an artificial model of TBG in order to show that fragile topology alone does not obstruct a gapped symmetric insulator \cite{turnerGappingFragileTopological2022a}. We likewise observe a gap between our quasiparticle peaks at $\nu \neq 0$. 

The light effective mass (which can be arbitrarily small) of a bound state built entirely out of heavy quasiparticles is thus a unique striking feature of our trions. It originates from the underexplored combination of Mott physics and band topology, two paradigmatic areas of condensed matter physics. We are not aware of any other three-particle bound states that have this property or origin. It is, however, worth comparing and contrasting with the quasiparticles of the heavy Fermi liquid. These are described as (higgsed) spinons or electrons dressed by a Kondo singlet, $\psi ~ c^\dag \sigma S$. These typically overlap with the electron, but if the singlet wavefunction has non-trivial angular momentum, the hybridization can be enforced to be zero at certain high-symmetry points \cite{ghaemiAngledependentQuasiparticleWeights2008}. There is an analogy here with the Dirac trion, which is topologically and symmetry protected from hybridizing with the electron at a single point and can be described as a higgsed ancilla fermion in which the condensate vanishes at that point. Despite this high level analogy, the excitations themselves are very different. The Dirac trion is light, composed entirely out of the most localized part of a band, topologically orthogonal to the electron, whereas the heavy fermion quasiparticle is heavy and combines a conduction electron with the spin flip of a second species.

    Here, we have focused on an effective model of TBG where we restrict to the active flat bands, while remote bands are excluded. This provides the minimal setting to explore TBG, and is consistent with existing estimates of energy scales \cite{ledwith2024nonlocal} and experiments measuring the spectral features   \cite{xieSpectroscopicSignaturesManybody2019a, YazdaniCascade, Nuckolls_2023, EvaAndreiSTM2019, STMNadjPerge, BandFlattening,IlaniCascade,saitoIsospinPomeranchukEffect2021,bocarsly2024imagingcoulombinteractionsmigrating,shahalnewexpt}. However, a different approach is to include mixing with remote bands to define local orbitals \cite{hauleMottsemiconductingStateMagic2019b,Bascones2019, Bascones2020, Bascones2023,song2022TBGTHF,calugaru2023TBGTHF2,yu2023TTGTHF,HuRKKY,herzogarbeitman2024topologicalheavyfermionprinciple, TBGKondo, THFKondo, TBGKondoSDS, SongLian, lau2024topologicalmixedvalencemodel, ValentiDMFT} and it is worthwhile to see how the Dirac trions can be approached from that starting point. For concreteness, let us consider the  topological heavy fermion (THF) model \cite{song2022TBGTHF, calugaru2023TBGTHF2,yu2023TTGTHF,HuRKKY,herzogarbeitman2024topologicalheavyfermionprinciple, TBGKondo, THFKondo, TBGKondoSDS, SongLian, lau2024topologicalmixedvalencemodel, ValentiDMFT} in two distinct limits: the projected limit $\gamma \gg U$ and the weak hybridization (or heavy fermion) limit $U \gg \gamma$. Here, $\gamma$ is the $c$-$f$ hybridization parameter in the THF language that determines the gap to the remote bands. As described in more detail  below, the trionic character of the $\Gamma$ point hole excitations at $\nu=-2$ is sensitive to the $\gamma/U$ ratio. In particular, it vanishes in the heavy fermion limit $\gamma/U \rightarrow 0$.
    
    In the heavy fermion limit, the two flat bands and four remote bands (per spin per valley here and below) are together best thought of as two topologically trivial AA-localized $f$ bands and four delocalized $c$ electron bands (two cones) whose $\Gamma$ wavefunctions have a node at the AA sites \cite{song2022TBGTHF}. The $f$'s are Mott-gapped to $\pm U$ and hybridize at order $\gamma/U \ll 1$ with the $\Gamma$-point $c$'s. In contrast, in the projected limit, the $\Gamma$-point $f$'s and half the $c$'s, which we refer to as $c^{(R)}$, are gapped to $\pm \gamma$ and become remote bands. The other two $c$'s', which we refer to as $c^{(F)}$ make up the $\Gamma$ points of the flat bands. The HF limit then has two Dirac cones, in which $c^{(F)}$ and $c^{(R)}$ define the two 'sublattices', while the projected limit also has two Dirac cones except with $c^{(R)}$ replaced by the Dirac trion. This suggests that the low energy $\Gamma$ excitation orthogonal to $c^{(F)}$ has the form 
	\begin{equation}
	\tilde{F}_\Gamma= u F_\Gamma + v c^{(R)}_\Gamma + w f_\Gamma,
        \label{eq:trion_cfermion}
	\end{equation}
    In the limit of vanishing hybridization, $\gamma/U \to 0$, we have $u \to 0$, $v \to 1$, and $w \to 0$, which means that $\tilde F_\bk$ reduces to $c^{(R)}_\bk$ and it does not have any trionic character. On the other hand, in the projected limit, $\gamma/U \to \infty$, we get $u \to 1$, $v \to 0$, and $w \to 0$ such that $\tilde F$ is purely trionic in nature. The recent ancilla calculations \cite{zhao2025ancilla}when applied to the THF model support \eqref{eq:trion_cfermion}. When applied to the IKS state, this is consistent with a recently proposed spectral function in the THF model \cite{herzog-arbeitmanKekuleSpiralOrder2025}, as well as the hole band edge of the $\nu = -2$ IKS state in the HF model that appear to be a mixture of $c^{(R)}_{\bk}$ and $F_\bk \equiv f_{\bk \pm q_s}$. 
    
    The previous discussion suggests that the trionic nature of the hole-like excitations of the $\nu = -2$ state is sensitive to the ratio $\gamma/U$. Recent QTM experiments \cite{shahalnewexpt} seem to indicate a splitting of $40 {\rm meV} \approx 2(U + E_{\rm str})$ at charge neutrality. Identifying the strain splitting as $2E_{\rm str} \approx 15$meV, this suggests a value of $U \lesssim 15 meV$. Note that this is consistent with our prior estimates based on earlier experiments \cite{ledwith2024nonlocal}. Meanwhile, the flat-remote gap is found to be $\gamma \gtrsim 60$meV. This places the parameters firmly within the projected limit and suggests only a small weight of $\tilde F$ on electrons: $Z_\Gamma = \abs{v}^2 + \abs{w}^2 \approx U^2/\gamma^2 \lesssim 0.1$.

	A Dirac trion as the light charge carrier could help solve a longstanding puzzle associated with the compressibility cascades \cite{CascadeShahal,CascadeYazdani}: the peaks in $d\mu/dn$ are shifted to the right of $\nu=1,2,3$ but not $\nu = 4$, which is the band insulator. The shift of the peaks relative to the integers is hard to explain from the perspective of zero temperature correlated insulators. In contrast, these shifts appear naturally in the limit of weak $c$-$f$ hybridization $\gamma \ll U$. Essentially one ``overdopes" the $c$-electrons relative to each integer due to the large Hubbard $U$ cost of doping states localized at AA. However, this picture suggests this overdoping would happen at \emph{all} integers, including $\nu = \pm 4$~\cite{HuRKKY,ledwith2024nonlocal}. This implies the absence of full filling gaps at $\nu = \pm 4$ in the limit $\gamma \ll U$ which is inconsistent with experiment. On the other hand, this issue can be evaded if the light particles are trions. To see this, note that trions have a finite density of states in the thermal state at $\nu = 1,2,3$, which can shift the $d\mu/dn$ peak by an $O(s^2)$ amount to the right. On the other hand, at $\nu \to +4$, we approach a mean-field state (even for normal state temperatures) where the trion spectral weight vanishes. Thus, the shift will vanish at $\nu = \pm 4$, where all the charge is converted into the electrons of the band insulator. 
    
    An important future work will be to verify this cascade scenario through computing $O(s^2)$ corrections to $dn/d\mu$. Such a calculation would also clarify the compressibility and transport properties of the Mott states at non-integer filling. We emphasize that the gap between the quasiparticle peaks we present does not imply a hard gap in compressibility and transport. A useful example for this point is the one-dimensional Hubbard model: in the limit $U \gg t $ there is a $2U$ gap between quasiparticle peaks but no charge gap for any $t > 0$ \cite{liebAbsenceMottTransition1968,liebOnedimensionalHubbardModel2003}. 
    
    Another important next step is to compute trion correlations in numericsl. Quantum Monte Carlo at intermediate temperatures at charge neutrality~\cite{hofmannFermionicMonteCarlo2022,MengQMCSemimetal} are consistent with the Mott semimetal electron spectral function. DMFT simulations at $U \sim \gamma$\cite{ValentiDMFT,bascones2025} have various features in common as well; it would be interesting to perform these at the now understood value $\gamma/U \gtrsim 4$ \cite{shahalnewexpt}. Sign-problem-free QMC at charge neutrality could provide an independent unambiguous verification of the trion and study its fate outside our analytical controlled setting, while DMFT simulations could provide an independent check on our trion cascade scenario.

	We now briefly comment on the implications for superconductivity and the TBG normal state. First we point out that as more experiments have been performed, the constraints on standard theories of superconductivity in TBG have become strongly restrictive. The superconductor forms adjacent to a correlated insulator and its normal high-temperature state is a strange metal\cite{PabloLinearT,YoungLinearT,EfetovSM}, $T_c$ continuously drops to zero upon approaching a van Hove singularity \cite{haoElectricFieldTunable2021,parkTunableStronglyCoupled2021}, the superconductor strongly violates the BCS Pauli limit \cite{caoPaulilimitViolationReentrant2021}, and features pseudogap spectra\cite{jiangChargeOrderBroken2019,ohEvidenceUnconventionalSuperconductivity2021,choiInteractiondrivenBandFlattening2021}. Particularly striking and puzzling is the fact that the superconductor appears to be nodal and phase-fluctuation-limited, with a zero-temperature non-monotonic superfluid stiffness that tracks $T_c$ \cite{banerjeeSuperfluidStiffnessTwisted2024}, while \emph{also} having a small superconducting tunneling gap that tracks $T_c$\cite{parkSimultaneousTransportTunneling2025}. To our knowledge, neither a conventional BCS, BEC, or intermediate crossover phase can capture these almost contradictory phenomena. We are, however, excited about the ability of Dirac trions to shed light on some of these questions. It is particularly worth noting that while the trions have a significant ``on-site" aspect to them, and could naturally attempt to form a BEC, they also contain topology-induced tails that extend over a long distance $\sim 1/s$. This dual nature of the trion could then be relevant for the corresponding mysterious dual-nature superconductivity. 

	A key step towards establishing the trion as the carrier that pairs and condenses will be to excite it. The momentum-conserving tunneling of the quantum twisting microscope \cite{QTM, shahalnewexpt} appears to not show any $\Gamma$ excitation on the hole (electron) side for $\nu < 0$ ($\nu > 0$). This is consistent with our picture: in the projected limit the $\Gamma$-point trion has no overlap with the electron and we estimated that corrections to this will introduce a $\lesssim 10\%$ electron weight.

	More broadly, our work suggests that there could be a multitude of composite particle states waiting to be uncovered in bands with strongly $\bk$-space quantum geometric inhomoegeneity. This is especially the case in topological bands far from Landau levels, a rapidly growing setting within moiré materials. Such bands necessarily have sharply distinct wavefunctions at some of their high symmetry points becuase the angular momenta for, e.g., $C_{3z}$ symmetry must sum to the Chern number modulo three. Then, many-body excitations symmetrically composed from one part of the band (the non-$\Gamma$ part in TBG) are naturally orthogonal to the wavefunction with distinct crystalline symmetry (the $\Gamma$ point). This is particularly natural when certain wavefunctions are more strongly interacting than others. Even when the initial crystal symmetry is dropped, we have seen that the vanishing overlap is topologically protected. In contrast, the lack of momentum space structure in typical topologically trivial bands and Landau levels means that orthogonal composite excitations occur through obtaining distinct internal symmetry quantum numbers or fractionalizing. We argue that these insights together with our controlled analytical techniques open the door to new kinds of exotic correlated physics that make use of strong interactions beyond electrons.

\section*{Acknowledgements}

We are grateful to Shahal Ilani for discussions regarding his experimental data for TBG spectral functions. We thank Qingchen Li, Pavel Nosov, and Ophelia Sommer for helpful discussions and related collaborations.  
A.V. is supported
by a Simons Investigator award, the Simons Collaboration on Ultra-Quantum Matter,
which is a grant from the Simons Foundation (651440,
A.V.)  and by NSF-DMR
2220703. EK is supported by an NSF CAREER award No. DMR-2441781.

\widetext
\appendix

\section{Derivation of the Schwinger-Dyson equations}

In this appendix we derive the SDEs quoted in the main text. To be self-contained, we restate the band projected Hamiltonian
\begin{equation}
  \H = \sum_\bk c^\dag_\bk h^{\rm sp} c_\bk + \frac{1}{2A}\sum_\bq V_\bq \delta \rho_\bq \delta \rho_{-\bq},
\end{equation}
where we refer to the second term as $\H_{\rm int}$. 

We begin with the imaginary time Green's function
\begin{equation}
  G^{cd}_{\bk \tau} = -\langle \T c_{\bk \tau} d^\dag_\bk\rangle.
  \label{eq:greenfndefn}
\end{equation}
Here $d^\dag_{\bk}$ is an arbitrary fermion operator, $\langle A \rangle = \frac{\Tr e^{-\beta \H} A}{\Tr e^{-\beta \H}}$ is the thermal expectation value, and all operators are propagated in imaginary time as $A_\tau = e^{\tau \H} A e^{-\tau \H}$. The time ordering $\langle \T A_\tau B_{\tau'}\rangle = \Theta(\tau-\tau')\langle A_\tau B_{\tau'}\rangle \pm \Theta(\tau'-\tau)\langle B_{\tau'} A_\tau\rangle$ where $+(-)$ corresponds to $A,B$ bosonic (fermionic). Since $c,d^\dag$ are fermionic, $G^{cd}_{\bk \tau+\beta} = -G^{cd}_{\bk \tau}$; we will later use this to Fourier transform to Matsubara frequencies $\omega_n = 2\pi (n+\frac{1}{2}) \beta^{-1}$. As usual the Green's function should be understood as a matrix in band indices. We have suppressed, and will continue to suppress, band indices $c^\dag_{\bk \alpha} ,d^\dag_{\bk \alpha}\to c^\dag_{\bk},d^\dag_{\bk}$ since the matrix structure will not play a significant role.

Differentiating \eqref{eq:greenfndefn} leads to
\begin{equation}
  -\partial_\tau G^{cd}_{\bk \tau} = \delta(\tau) \langle\{c_{\bk}, d^\dag_\bk\}\rangle + \langle \T [\H, c_{\bk\tau}] d^\dag_\bk \rangle.
  \label{eq:diffgreenone}
\end{equation}
We now write
\begin{equation}
    -[\H, c_{\bk \tau}] = -e^{\tau \H} [\H , c_{\bk}] e^{- \tau \H} = h^\sp c_{\bk \tau} + O_{\bk \tau}, \qquad 
    O_{\bk} = -[H_{\rm int}, c_{\bk}] = \frac{1}{2A} \sum_\bq V_\bq \Lambda_{\bk,\bk+\bq}\{ c_{\bk + \bq}, \delta \rho_{-\bq} \}
\end{equation}
Inserting back into \eqref{eq:diffgreenone} leads to
\begin{equation}
\begin{aligned}
  -\partial_\tau G^{cd}_{\bk \tau} & = \delta(\tau) \langle\{c_{\bk}, d^\dag_\bk\}\rangle + h^\sp _\bk G^{cd}_{\bk \tau} +  G^{Od}_{\bk \tau} 
  \label{eq:diffgreentwo}
  \end{aligned}
\end{equation}

We now Fourier transform to Matsubara frequencies $\omega_n = 2\pi (n+\frac{1}{2})/\beta$ via $G_{\bk n} = \int_{0}^\beta d\tau \,e^{i \omega_n \tau} G_{\bk \tau}$. We obtain
\begin{equation}
\begin{aligned}
  (i\omega_n - h^\sp_\bk)G^{cd}_{\bk n} & =  \langle\{c_{\bk}, d^\dag_\bk\}\rangle  +  G^{Od}_{\bk n} 
  \label{eq:diffgreenthree}
  \end{aligned}
\end{equation}
Substituting $d\to c$ above leads to
\begin{equation}
\begin{aligned}
  G_{\bk} & = \frac{1}{i\omega_n - h^\sp_\bk - \Sigma_{\bk n}}, \qquad \Sigma_{\bk,n} =G^{Oc}_{\bk n} G^{-1}_{\bk n} = G^{-1}_{\bk n}G^{cO}_{\bk n},
  \label{eq:electronGsupp}
  \end{aligned}
\end{equation}
where we identified the self energy $\Sigma_{\bk n}$ and used the shorthand $G = G^{cc}$ for the electron-electron Green's function. The second expression for $\Sigma_{\bk n}$ comes from taking the Hermitian conjugate of $G$ or applying analogous steps to $-\langle \T c_\bk c_{\bk -\tau}\rangle$.

To go further we introduce the 1PI correlators. We define $[G_{\bk,n}^{O,d}]_{\rm 1PI} = G_{\bk n}^{O,d} - [G_{\bk,n}^{O,d}]_{\rm 1PR}$. Here $[G_{\bk,n}^{O,d}]_{\rm 1PR} = G_{\bk n}^{Oc}G^{-1}_{\bk n}G^{cd}_{\bk n}$ consists of the contributions to $G_{\bk n}^{O,d}$ that involve a single electron excitation as an intermediate state. As a result, $[G_{\bk,n}^{O,d}]_{\rm 1PI}$ only receives contributions from intermediate states with multiparticle excitations. 

From \eqref{eq:electronGsupp} we identify $[G_{\bk n}^{O,d}]_{\rm 1PR} = \Sigma_{\bk n} G^{cd}_{\bk n}$ such that \eqref{eq:diffgreenthree} can be written as
\begin{equation}
    G^{-1}_{\bk n} G^{cd}_{\bk n} =  \langle\{c_{\bk}, d^\dag_\bk\}\rangle  +  G^{Od}_{\bk n}.
\end{equation}
This is the equation we use repeatedly in the main text.

We now compute the anticommutators $\langle\{c_{\bk}, d^\dag_\bk\}\rangle$ where $d^\dag_\bk =O^\dag_\bk, F^\dag_\bk$. In both cases we use the identity $\{A,\{B,C\}\} = \{\{A,B\},C\}-[B,[A,C]]$. We start with $O^\dag_\bk$ and evaluate
\begin{equation}
  \begin{aligned}
    \langle \{c_{\bk,\alpha}, \{c^\dag_{\bk+\bq,{\beta}}, \delta \rho_\bq\} \} \rangle & = \langle\{\{c_{\bk \alpha}, c^\dag_{\bk+\bq \beta}\}, \delta \rho_\bq\}\rangle - \langle[c^\dag_{\bk+\bq \beta},[c_{\bk \alpha}, \delta \rho_\bq]]\rangle\\
    & = 2\delta_{\alpha \beta} \sum_\bG \delta_{\bq,\bG}  \langle  \delta \rho_{\bG} \rangle -[\Lambda_{\bk,\bk+\bq}Q_{\bk+\bq}]_{\alpha \beta}.  \\
\end{aligned}
  \label{eq:anticommutator}
\end{equation}
such that
\begin{equation}
    \langle \{ c_\bk, O^\dag_\bk \} \rangle = \frac{1}{A}\sum_\bG V_\bG \Lambda_{\bk,\bk+\bG} \langle \delta \rho_\bG \rangle - \frac{1}{2A}\sum_\bq V_\bq \Lambda_{\bk,\bk+\bq} Q_{\bk + \bq} \Lambda_{\bk+\bq,\bk} = h^\mf_\bk
\end{equation}

For the trion, we recall the expression $F^\dag_{\bk \alpha} = \frac{1}{\sqrt{N}} \sum_\bR e^{i \bk \cdot \bR} \{c^\dag_{\bR}, \delta n_\bR \} - [c^\dag_\bk A^\dag_\bk]_\alpha $ where $A_\bk$ is chosen to make the anticommutator with the electron vanish and $c^\dag_{\bR \alpha} = \frac{1}{\sqrt{N}} \sum_\bR \chi_{\bk \alpha} c^\dag_{\bk \alpha} e^{-i\bk \cdot \bR} $. Here and below there are no sums over repeated indices and we label $\chi$ with $\alpha$; it depends on the Chern sector $\gamma$ that $\alpha$ indexes alongside spin and valley. We begin by computing the anticommutator with the first term in $F^\dag_{\bk \beta}$:
\begin{equation}
\begin{aligned}
    \frac{1}{\sqrt{N}}\sum_\bR e^{i \bk \cdot \bR} \langle \{c_{\bk,\alpha}, \{c^\dag_{\bR,{\beta}}, \delta n_\bR\} \} \rangle & = \frac{1}{\sqrt{N}}\sum_\bR e^{i \bk \cdot \bR} \left(\langle\{\{c_{\bk \alpha}, c^\dag_{\bR\beta}\}, \delta n_\bR\}\rangle - \langle[c^\dag_{\bR \beta},[c_{\bk \alpha}, \delta n_\bR]]\rangle \right) \\
    & = \ov{f}_{\bk \beta} \left(2\delta_{\alpha \beta}\langle \delta n_\bR \rangle - \frac{1}{N}\sum_\bR \langle [c^\dag_{\bR \beta}, c_{\bR \alpha}]  \rangle\right). \\
    & = \ov{f}_{\bk \beta}\left(2\delta_{\alpha \beta}\langle \delta n_\bR \rangle - \frac{1}{N}\sum_\bk \abs{\chi_\bk}^2 Q_{\bk \alpha \beta} \right) \\
    & := [A_{\bk}^\dag]_{\alpha \beta}.
    \end{aligned}
\end{equation}
We therefore have
\begin{equation}
    \{c_{\bk \alpha}, F^\dag_{\bk \beta} \} = [A_{\bk}^\dag]_{\alpha \beta} - \{ c_{\bk \alpha}, [c^\dag_\bk A^\dag_\bk]_\beta \} = 0,
\end{equation}
for the $A_\bk$ that we claimed in the main text.

\end{document}